\documentclass[aps, prb, reprint, showpacs, superscriptaddress]{revtex4-1}

\usepackage{graphicx}
\usepackage{dcolumn}
\usepackage{amsmath}
\usepackage{subfigure}
\usepackage{color}
\usepackage{float}
\usepackage[normalem]{ulem}

\begin{document}

\title{Textured heterogeneity in square artificial spin ice }

\author{J. C. T Lee}
\thanks{These authors contributed equally to this work.}
\affiliation{Advanced Light Source, Lawrence Berkeley National Laboratory, Berkeley, CA 94720-8229, USA}
\affiliation{Department of Physics, University of Oregon, Eugene, OR 97403-1274, USA}

\author{S. K. Mishra}
\thanks{These authors contributed equally to this work.}
\affiliation{Advanced Light Source, Lawrence Berkeley National Laboratory, Berkeley, CA 94720-8229, USA}

\author{V. S. Bhat} 
\affiliation{Department of Physics and Center for Advanced Materials, University of Kentucky, Lexington, KY 40506-0055, USA}

\author{R. Streubel} 
\affiliation{Materials Sciences Division, Lawrence Berkeley National Laboratory, Berkeley, CA 94720-8229, USA}

\author{B. Farmer} 
\affiliation{Department of Physics and Center for Advanced Materials, University of Kentucky, Lexington, KY 40506-0055, USA}

\author{X. Shi}
\affiliation{Advanced Light Source, Lawrence Berkeley National Laboratory, Berkeley, CA 94720-8229, USA}
\affiliation{Department of Physics, University of Oregon, Eugene, OR 97403-1274, USA}

\author{L. E. De Long}
\affiliation{Department of Physics and Center for Advanced Materials, University of Kentucky, Lexington, KY 40506-0055, USA}

\author{I. McNulty}
\affiliation{Center for Nanoscale Materials, Argonne National Laboratory, Argonne, IL 60439, USA}

\author{P. Fischer} 
\affiliation{Materials Sciences Division, Lawrence Berkeley National Laboratory, Berkeley, CA 94720-8229, USA}
\affiliation{Department of Physics, University of California, Santa Cruz, CA 95064, USA}

\author{S. D. Kevan} 
\affiliation{Advanced Light Source, Lawrence Berkeley National Laboratory, Berkeley, CA 94720-8229, USA}
\affiliation{Department of Physics, University of Oregon, Eugene, OR 97403-1274, USA}

\author{S. Roy}
\email[Corresponding author: ]{sroy@lbl.gov}
\affiliation{Advanced Light Source, Lawrence Berkeley National Laboratory, Berkeley, CA 94720-8229, USA}

\date{\today}
 
\begin{abstract}

We report evidence of spontaneous formation of a heterogeneous network of superdomains in two-dimensional square artificial spin ice nanostructures in externally applied magnetic fields.
The magnetic heterogeneity is locally disordered but has a zig-zag texture at longer length scales.
Resonant coherent soft-x-ray scattering off such textures give rise to unique internal structure in Bragg peaks.
Our result shows that the macroscopic magnetic texture is derived from the microscopic structure of the Dirac strings.

\end{abstract}

\maketitle
Artificial spin ice (ASI) systems are magnetic nanostructures whose magnetization textures mimic the frustrated hydrogen bonding networks observed in water ice.\cite{Wang, Nisoli, Pauling}
ASI are most commonly periodic arrays of identical, elongated thin film islands whose shape anisotropy forces their magnetizations to align along their long axes, which gives them an Ising spin character.
Geometrical constraints dictate the interactions between the Ising spins. 
In square ASI, for example, the arrangement of mutually perpendicular nanomagnets over a square lattice causes an asymmetry in the inter-island interactions that, in principle, favor an antiferromagnetic ground state.
However, high-energy barriers to magnetization switching of individual islands oppose their thermalization at room temperature, and the ground state is attained only in limited areas of the ASI.\cite{Kapaklis,Arnalds}.

Magnetic excitations of ASI are analogous to those of natural (atomic scale) spin ice\cite{Fennell1, Fennell2, Morgan, Mol}, and are quasiparticle-like magnetic point charges that reflect the behavior of hypothetical magnetic monopoles.\cite{Jaubert}
Pairs of opposite-polarity charges are connected by a chain of flipped spins, or ``Dirac strings'', which result from a chain of successive Ising spin-flips against a background spin texture.\cite{Castelnovo, Mol}

The formation and propagation of Dirac strings during the magnetization reversal is particularly interesting as a general process among ASI. 
In Kagome ASI, Mengotti \textit{et al.}\cite{Mengotti} found that reversals occur via nucleation, followed by avalanches of magnetic charges.
In contrast, square ASI with large lattice constants form pairs of opposite charges that remain near randomly distributed vertices, (\textit{i.e.} junctions between the nanoislands) instead of becoming itinerant.\cite{Phatak-NJP}
When such charges do move, the resulting Dirac strings prefer to form closed loops rather than open-ended chains.\cite{Morgan}
These studies show that square ASI spin textures are heterogeneous and form structures that span many vertex sites.
In order to distinguish between domains that occasionally form within each nanoisland from the areas of uniform ASI spin texture, we denote these mesoscale magnetic structures \textit{superdomains}.
While concepts of heterogeneity are central to the global understanding of spin ice physics, they have not been addressed in any detail.
It is important to understand the evolution of magnetic heterogeneity on a macroscopic length scale, and how energetic considerations in both micro- and macroscopic length scales influence formation of superdomains, which will affect nucleation and propagation of magnetic charges.

Herein, we demonstrate that the constructive and destructive interference of a coherent x-ray beam scattered from a square ASI results in Bragg peaks whose internal structure is highly sensitive to heterogeneity in the magnetic texture. 
By analyzing Bragg peak profiles, we extract detailed information of the various spin textures within a macroscopic area. 
We found that after taking a square ASI through multiple hysteresis loops, a partially magnetized two dimensional (2D) square ASI develops a heterogeneous patchwork of areas, each of which can extend over tens of lattice sites, with the same vertex states (\textit{e.g.}, ``two-in/two-out'' T2 textures) throughout.
Although locally heterogeneous, over large length scales the disordered superdomains have a zig-zag texture that is akin to the zig-zag shape of Dirac strings propagating along a diagonal (e.g., [11]) direction of square ASI. 
The heterogeneity is stable under application of moderate applied fields, and the ASI ``remembers'' the arrangement of superdomains from previous hysteresis loops, which is evidence for magnetic pinning sites.\cite{GilbertPRB}

\begin{figure}
\includegraphics[width= 8.6 cm]{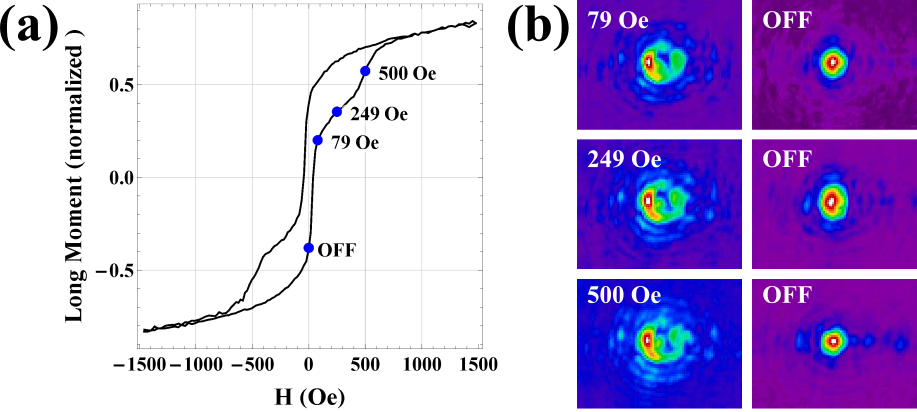}
\caption{
(a) Hysteresis loop of the square ASI sample. 
The magnetic field was applied along the [10] direction of the square ASI and the magnetization (``Long Moment'') was normalized to the saturation value $M_S$. 
Blue dots show field points of 79, 249 and 500 Oe, at which x-ray measurements were done.  
(b) Bragg peaks in resonant coherent x-ray diffraction patterns from the square ASI. 
Bragg peaks split into a rough ring shape when applied magnetic field is on during a hysteresis loop (left column) and become approximately circular (due to pinhole) when the field is off (right column). 
This ring shape persists till $\mu_0H \approx$ 80 mT, though it subtly changes at higher fields.   
}
\label{F1}
\end{figure}

Our samples were 2D square ASI fabricated from Permalloy (Ni$_{0.81} $Fe$_{0.19}$) deposited on a Si wafer using electron beam deposition.   
The nanoislands were fabricated using electron beam lithography, yielding thicknesses \emph{t} = 25 nm, widths \emph{w} = 50 nm, and lengths \emph{$\ell$} = 150 nm, and square ASI lattice constant \emph{d} = 300 nm.
Sample fabrication details are given in the Appendix.
Resonant coherent soft x-ray scattering was performed at a low incidence angle of 9$^{\circ}$ at Beamline 12.0.2.2 at the Advanced Light Source, Lawrence Berkeley National Laboratory. 
The coherent x-ray beam was obtained by placing a 10 $\mu$m pinhole at the monochromator focus, located 5 mm upstream from the ASI.
Our coherent $\sigma$ polarized incident x-ray beam was resonantly tuned to the Fe  L$_3$ edge (707 eV) to enhance the magnetic contrast. 
On resonance, $\sigma$ polarized soft x-rays are sensitive to the component of magnetization along the beam direction (\textit{i.e.}, $\vec{k_0}$$\parallel$$\vec{M}$) \cite{Kortright}. 
Taking all of this into account, we probed $\approx$5,580 nanoislands whose long axes were parallel both with the applied magnetic field and incident beam.
We applied magnetic fields along the beam direction to manipulate the ASI magnetization and used charge-coupled device camera, 0.5 m downstream of the ASI, to record diffraction patterns at several applied magnetic fields. 

Fig. \ref{F1}(a) shows the DC magnetic hysteresis loop of the sample with magnetic field applied along the [10] direction of the square spin ice unit cell. 
The blue dots mark the field positions where the resonant coherent x-ray diffraction data was taken.
In order to study the field evolution of magnetic scattering, we field cycled the sample several times (about 5 times between $\pm {0.25}$ T) through the hysteresis loop before recording any diffraction data.
We then applied a saturating field opposite to the beam propagation direction, and then applied the fields of 10 mT $ \leq \mu_0$H$_z$ $\leq$ 80 mT.
Every field condition measurement was followed by a zero field measurement. 

In zero field conditions, we obtained symmetric Bragg peaks surrounded by diffuse Airy fringes caused by diffraction from the circular pinhole.
The intensity pattern is remarkably different from those acquired in non-zero magnetic field (Figure \ref{F1}(b), left panel). 
``Doughnut''-shaped Bragg peaks form with intensity minima at their centers surrounded by non-uniform circular rings of intensity that features two strong intensity regions along the scattering plane, connected by weaker arcs of intensity.
Interestingly, on switching off the magnetic field, we observe symmetric Bragg peaks without any discernible ring-like structure ((Figure \ref{F1}(b), (Right panel). 
Since the Bragg peaks develop internal structure with the application of magnetic fields, changes as the field is increased, and returns to a symmetric shape when the field in taken away, we conclude that the resonant scattering effect shown in Fig. \ref{F1}(b)(left panel) is magnetic in origin. 

\begin{figure}
\includegraphics[width= 8.6 cm]{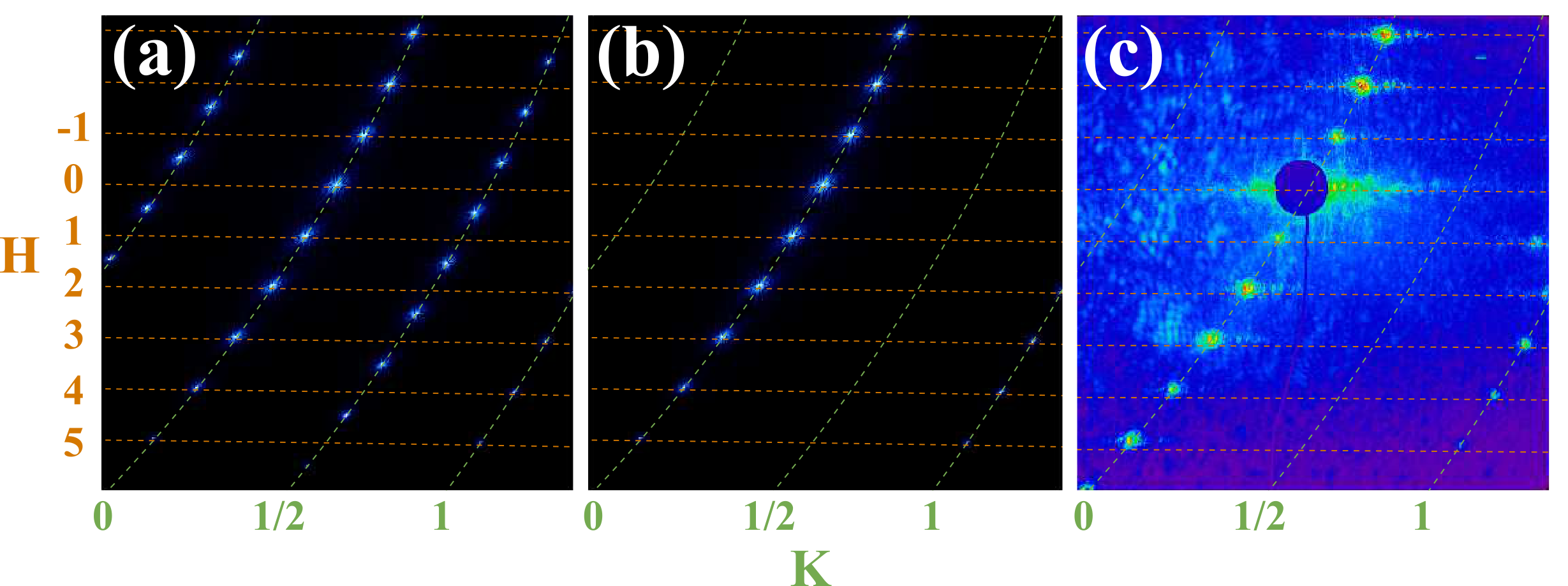}
\caption{ 
Ideal resonant magnetic x-ray scattering from T1 and T2 vertex lattices vs. data. 
(a) Characteristic scattering from a lattice of T1 vertices. 
The antiferromagnetic arrangement of spins cause peaks at half-integer Miller index positions (\textit{e.g.}, peaks along dashed $K = 1/2$ line). 
(b) ``Half-order'' peaks are absent in scattering from a T2 vertex lattice.  
(c) Zero-field diffraction data. 
Lack of half-order peaks indicates that no long range T1-order exists in the sample. 
}
\label{F2} 
\end{figure}

In a 2D square ASI the four islands surrounding a vertex site can adopt $2^4=16$ distinct local spin configurations which can be assigned one of four energy states: T1, T2, T3 and T4. 
(See Appendix.)
In order to understand the mix of vertex states that could give rise to the observed scattering patterns, we performed model scattering calculations for cases where the spin ice is populated by either the lowest-energy T1 or next-lowest-energy T2 vertices (see Figure \ref{F2}).
Lattices entirely composed of T2 vertices result in Bragg peaks that precisely overlap with lattice Bragg peaks (Fig. \ref{F2}(b)), which are denoted by integer Miller $(H, K)$ indices.
A pure T1 lattice forms additional peaks at half integer Miller index positions,\textit{e.g.}, peaks lying on the green dashed lines and between yellow lines in Fig. \ref{F2}(a), consistent with the doubled unit cell of the T1 state, which is also the ideal ASI ground state.
Since our data (Fig. \ref{F2}(c)) does not exhibit any scattering at half-integer positions, we conclude that no long range T1 order exists in the sample.  
This is expected as the sample was subjected to several magnetic field cycles between maximum fields of $\pm 0.25$ T, which would have introduced a heterogeneous mixture of mainly T2 vertex states.\cite{Farhan}

\begin{figure}
\includegraphics[width= 8.6 cm]{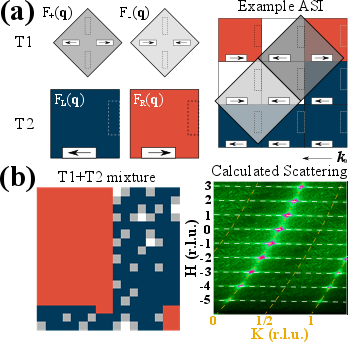}
\caption{ 
(a) Schematics of the unit cells used to calculate the magnetic structure factor of model ASI.
The scattering structure factors of the unit cells are $F_+(\vec{q})$ and $F_-(\vec{q})$ respectively for T1 unit cells with two-in or two-out arrow patterns; and $F_L(\vec{q})$ and $F_R(\vec{q})$ for T2 unit cells with left and right pointing arrows.
An example of how the unit cells build up an ASI is shown on the right.
(b) Calculated ASI diffraction patterns used to establish both an upper limit on the number of T1 vertices and their distribution in the ASI. 
If the T1 vertices do not form long-range order, they cause weak diffuse scattering at half-integer $(H, K)$, as in the calculated scattering pattern.
}
\label{F3}
\end{figure}

We defined a set of magnetic unit cells, shown in Fig. \ref{F3}, to construct model ASI spin textures and simulate resonant x-ray scattering patterns from magnetically heterogeneous ASI.
Figure \ref{F3}(a) schematically shows the unit cells and the nanoislands (containing arrows) that resonant magnetic x-ray scattering can distinguish in our scattering geometry.
The unit cells are divided into T1 and T2 types that describe square ASI when only one type of unit cell is present.   
T1/T2 vertex mixtures can also be made, as shown on the right of Fig. \ref{F3}(a).

The number of each unit cell type and their relative positions in the ASI determine the resonant magnetic x-ray scattering pattern.
Each unit cell type has a characteristic x-ray scattering structure that gives rise to a specific Bragg peak at a specific location in the reciprocal space.\cite{Note1}
ASI with only T1 unit cell structure factors $F_+$ or $F_-$ exhibit half-order scattering (Fig. \ref{F2}(a)), whereas T2 structure factors $F_L$ or $F_R$ show no half-order peaks (Fig. \ref{F2}(b)).
The distribution of T1 and T2 unit cells over the lattice significantly affects the x-ray scattering intensity which is determined by a sum of structure factors, weighted by position-dependent phases, over every lattice point $(i,j)$ of the ASI: 
$I(\vec{q}) \propto | \sum_{ i=0 }^{ N_x-1 } \sum_{ j=0 }^{ N_y-1 } F_{ij}(\vec{q}) \exp[ i \vec{q} \cdot \vec{r}_{ij}] | ^ 2$, where there are $N_x \times N_y$ lattice points. 
The constructive and destructive interference of terms with different structure factors can cause Bragg peaks to exhibit internal structure determined by the particular distribution of the unit cells. 
Coherent x-ray scattering detects these interference effects and reveals detailed information about the magnetic morphology and texture in a material.

We varied the spatial distribution and relative fraction of T1 and T2 unit cells in the ASI in our calculations to identify spin textures that replicated essential features of the data. 
For example, of the scenarios investigated, we found that T1 unit cells play no role in splitting the Bragg peak intensities. 
Figure \ref{F3}(b) illustrates the types of T1 (gray) and T2 (blue/red) unit cell mixtures we considered in our model ASI; a calculated diffraction pattern from one such mixture is shown on the right. 
To keep the half-order scattering intensity in our calculations below what would be the background level in our experimental data required that the T1 unit cells be sparsely distributed or only form a few small clusters containing no more than $\sim$ 50 unit cells.
No more than $\sim$ 10\% of the ASI lattice sites are occupied by T1 unit cells.

\begin{figure}
\includegraphics[width= 8.6 cm]{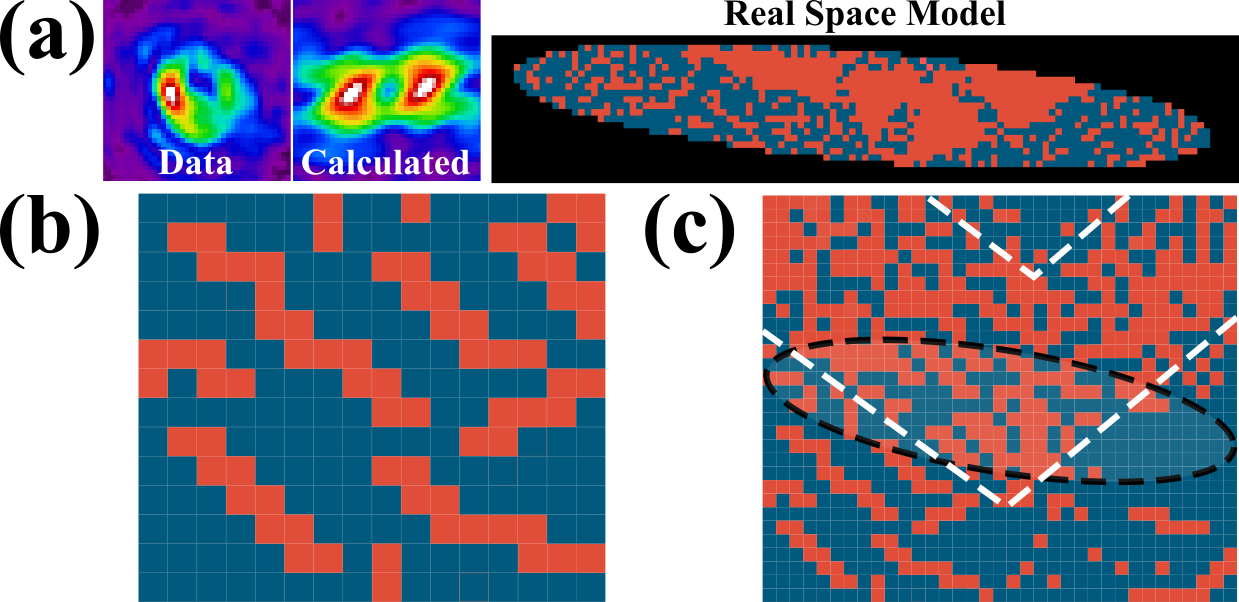}
\caption{ 
(a) Comparison between ``doughnut-hole'' Bragg peak data to Bragg peaks in calculated x-ray scattering pattern. The real space model on which the calculation is based is shown on the right.
(b) Schematic of Dirac strings forming out of the saturated state of an ASI. These strings prefer to move diagonally due to the symmetry of the square lattice.  
(c) Depiction of superdomains (bounded by white dashed lines) formed out of bundled Dirac strings. The dashed ellipse illustrates the area sampled by a x-ray beam. 
}
\label{F4}
\end{figure}

Given our signal is dominated by T2 unit cells, we then focused on the distributions of T2 unit cells that would yield the observed scattering pattern.
We found that random mixtures of small T2 states produce weak or no Bragg peaks.
As in the case of diffraction off crystals, ASI will give the strongest scattering when it has large areas of identical vertices/unit cells to provide constructive interference at the Bragg peaks.
In contrast, scattering amplitudes from ASI with many small ``crystallites'' of vertex order essentially add together incoherently (\textit{i.e.}, insignificant constructive interference). 
Therefore, the spin texture we observe must have large areas, or superdomains, of predominantly one T2 unit cell type (the unit cells are shown in Fig. \ref{F3}(a)).

The reproducible nature of the Bragg peak profiles over six consecutive hysteresis loops supports this conclusion.
This memory effect indicates quenched disorder in the ASI which promotes the formation of superdomains.
Based on changes in Bragg peak splitting shape and intensities, we estimate that at least 2.5\% of illuminated area of ASI changes from one hysteresis cycle to the next.
These differences in Bragg peak profile are slight, indicating that there is sufficient disorder in the magnetic texture to create superdomains containing hundreds of unit cells circumscribed by long, partially charged, boundaries.\cite{Reichardt}
Furthermore, this memory effect also implies the existence of a complex network of interacting magnetic defects, such as Dirac strings and magnetic charges.\cite{GilbertPRB}

Superdomain morphology strongly influences the overall shape and intensity of the scattered signal.
For example, antiphase domains in real crystals composed of unit cells of different structure factors can broaden Bragg peaks in particular directions in reciprocal space. \cite{warren} 
In the context of x-ray scattering from ASI, one superdomain (say, $+M$) can advance the phase of the scattered x-rays and the another adjacent one ($-M$) can retard it. 
The mismatched phases of the scattered light from the two domains will create a dark band going through the Bragg peak.
This Bragg peak splitting is particular to this magnetic texture, so seeing such a pattern is a strong indication that two large, oppositely polarized superdomains are being illuminated. 
Since a point defect in a crystal or a fork shaped structure gives rise to doughnut-shaped Bragg peaks, we examined various superdomain arrangements (see Appendix) that would resemble such a structure. 

Superdomains with walls that form an intersection or a convergence point is one such arrangement. 
The superdomain morphology and distribution that gave diffraction closest to the data is shown in Fig. \ref{F4}(a).
To account for experimental conditions in our real space model, we bounded the ASI by an ellipse to represent the beam footprint at 9$^{\circ}$ incidence.
The real space structure contains two different T2 superdomains each interspersed with a disordered background, and have diagonal walls that converge to form a cusp. 
The calculated diffraction pattern from the real space model, as shown in Fig. \ref{F4}(a), has distinct doughnut-like Bragg peaks with two strong intensity lobes separated by approximately the same reciprocal space distance as in the data.
The minima of the doughnut peaks arises from the presence of contrasting superdomains.
The doughnut shape, in particular, has its origin in the wedge-like superdomain morphology. 
This reveals an important clue regarding the mechanism directing superdomain morphology.

Starting from saturation, where the sample is Ising saturated, reversal begins by random spin flips giving rise to oppositely pointed T2 unit cells.
Islands in neighboring vertices flip and form areas dense with Dirac strings that traverse the diagonal easy directions of the square ice (see Fig. \ref{F4}(b)).
With increasing field, the Dirac strings grow in length and number. 
Their mutual repulsion promotes bundles of parallel Dirac strings to grow. 
These bundles ultimately form large T2 superdomains with net magnetization parallel to the applied field.

A magnetic field applied along [10] is equally likely to create Dirac strings growing along the easy directions [11] and [1$\bar{1}$].
Existing Dirac strings can also change their propagation directions from one easy axis to another at any given lattice site, causing cusp-like features in the chains of flipped spins. 
As bundles accumulate more Dirac strings with either differing growth directions or cusps, the growth directions of the bundles change.
Consequently, Dirac string bundles, and the T2 superdomains they engender, display morphologies with cusp-like features, as shown in Fig. \ref{F4}(c).
The formation of cusps on two different length scales indicates that the interactions between Dirac strings that control the superdomain morphology and heterogeneity are self-similar to inter-island interactions that control the course of meandering Dirac strings.
This implies that a scaling relationship between the two exists, a possibility we plan to further investigate.

Future experiments should be directed towards the topic of superdomain defects, memory effects, and their relationships to fundamental Dirac string excitation.
Topological defects, in the form of Buergers defects, can be introduced to ASI to create magnetic heterogeneity.\cite{Drisko}
Superdomains nucleated around the defects could form, each with a potentially different capacity to remember past states based on the quenched disorder, field history, and Buergers vector.
Coherent resonant x-ray scattering can characterize these superdomains and their persistence across magnetic field cycles. 

Generalizing to other frustrated systems, the important question is if  the properties of large areas of long-range order (\textit{e.g.}, superdomains) is determined by the fundamental properties of the excitations.
X-ray photon correlation spectroscopy or inelastic x-ray scattering can reveal the characteristic time and type of superdomain dynamics associated with memory effects.

\begin{acknowledgments}
J.C.TL., R.S., P.F., and S.D.K. acknowledge support by the Director, Office of Science, Office of Basic Energy Sciences, Materials Sciences and Engineering Division, of the U.S. Department of Energy (DOE), Contract No. DE-AC02-05-CH11231 within the Nonequilibrium Magnetic Materials Program (KC2204).
This research used resources of these DOE Office of Science User Facilities: the Advanced Light Source (DE-AC02-05CH11231); and the Advanced Photon Source (DE-AC02-06CH11357).
Work at the University of Oregon was partially supported by DOE, Office of Basic Energy Sciences, Materials Science and Engineering Division (DE-FG02-11ER46831). 
Research at the University of Kentucky was supported by DOE (DE-FG02-97ER45653) and the U.S. National Science Foundation (DMR-1506979).
\end{acknowledgments}

\appendix
\renewcommand{\thesection}{\Alph{section}}

\section{Magnetic vertices in square artificial spin ices}
\renewcommand{\thefigure}{\thesection} 

\begin{figure}[b]
\begin{center}
\includegraphics[width=8.75 cm]{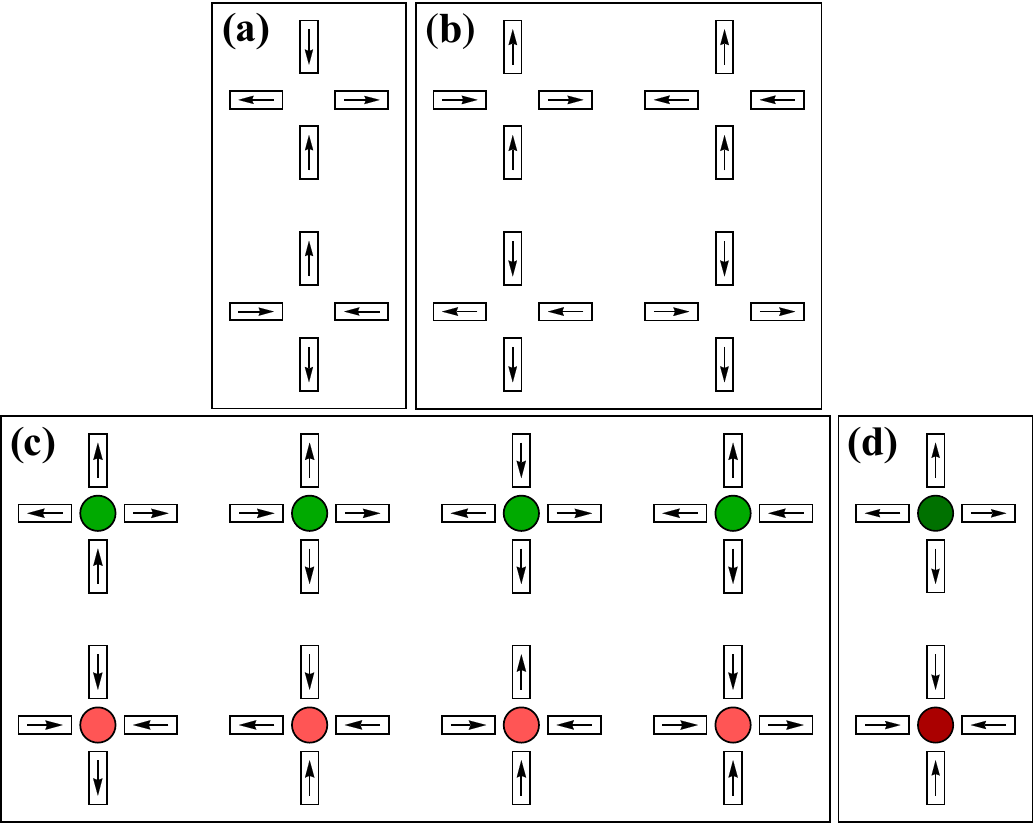}
\caption {
The Ising spin textures associated with (a) T1, (b) T2, (c) T3, and (d) T4 energy states of square ASIs. 
The two lowest energy configurations, T1 and T2, have two of four magnetic moments oriented toward the vertex and two oriented away from the vertex, analogous to the ``two in/two out'' ice rules for tetrahedral water ice. 
The two highest energy configurations, T3 and T4, display magnetic charges at their vertices, shown by green and red dots. 
}
\end{center}
\label{vertices}
\end{figure}

In a 2D square ASI the four islands that form a vertex can adopt 16 distinct local spin configurations which can be assigned to one of four energy states, called T1, T2, T3, and T4.
(See Fig. \ref{vertices}.) 
Analogous to the ``two-in/two-out'' ice rules for the atomic moments in a tetrahedral spin ice structure, T1 states, with one pair of opposing magnetic moments pointing inward and one pointing outward  from a vertex, have the lowest magnetostatic energy at zero applied field.
T2 states also have  two-in/two-out local spin textures but with pairs of opposing inward and outward moments. 
T2 states have higher energies than T1 states because of the inequivalent distance between the four nanobars which introduces asymmetry in the interaction energy between the four elements of a vertex.
States having three (T3) or four (T4) spins pointing in or out, respectively, exhibit higher magnetostatic energies and net magnetic flux (i.e., local magnetic monopoles).  
Generally, after multiple field cycling, 2D square ASIs attain states that are primarily mixtures of T1 and T2 configurations with sporadic T3 vertices.\cite{Farhan}

\section{Sample growth procedure}
\label{growth}

\begin{figure}[ht]
\begin{center}
\includegraphics[width=8.75 cm]{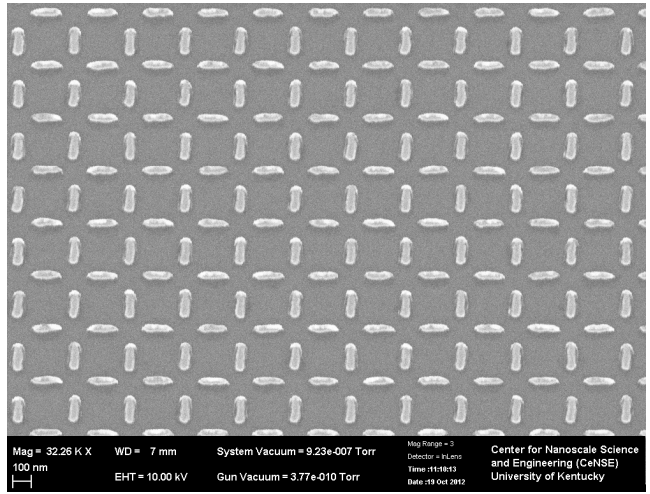}
\caption {Scanning electron micrograph of a square array of Permalloy dots of thickness \emph{t} = 25 nm, width \emph{w} = 50 nm, length \emph{$\ell$} = 150 nm, lattice constant \emph{a} = 300 nm, and total dimension of the array is 2 mm x 2 mm.}
\end{center}
\label{SEM}
\end{figure}

We have used electron beam lithography to pattern square lattices of Permalloy dots of thickness \emph{t} = 25 nm, width \emph{w} = 50 nm, length \textit{l} = 150 nm, and lattice constant \textit{a} = 300 nm. 
ZEP positive resist was spin-coated on a Si wafer prior to electron beam exposure. 
After the e-beam exposure and development, a Permalloy film of thickness 25 nm was then deposited using electron beam evaporation, with a base pressure of 10$^{-7}$ Torr. 
Final lift-off of resist was done using N-Methyl-2-pyrrolidone (NMP). 
Our sample had a 2 x 2 mm overall dimension.
A scanning electron micrograph of a portion of the sample is shown in Fig. \ref{SEM}.

\section{Calculation of magnetic structure factor}

The magnetic structure factor, being proportional to the scattered field, can be used to calculate the far field interference pattern.
We will explain how we arrived at expressions for the magnetic structure factors used in calculations that rendered Figures 3 and 4 of the main text.

Calculating the structure factors involves defining the repeating motif, or basis, decorating the square ASI lattice, then performing a Fourier transform of the nanoislands in the basis.\cite{warren}

\begin{figure}[ht]
\begin{center}
\includegraphics[width=8.75 cm]{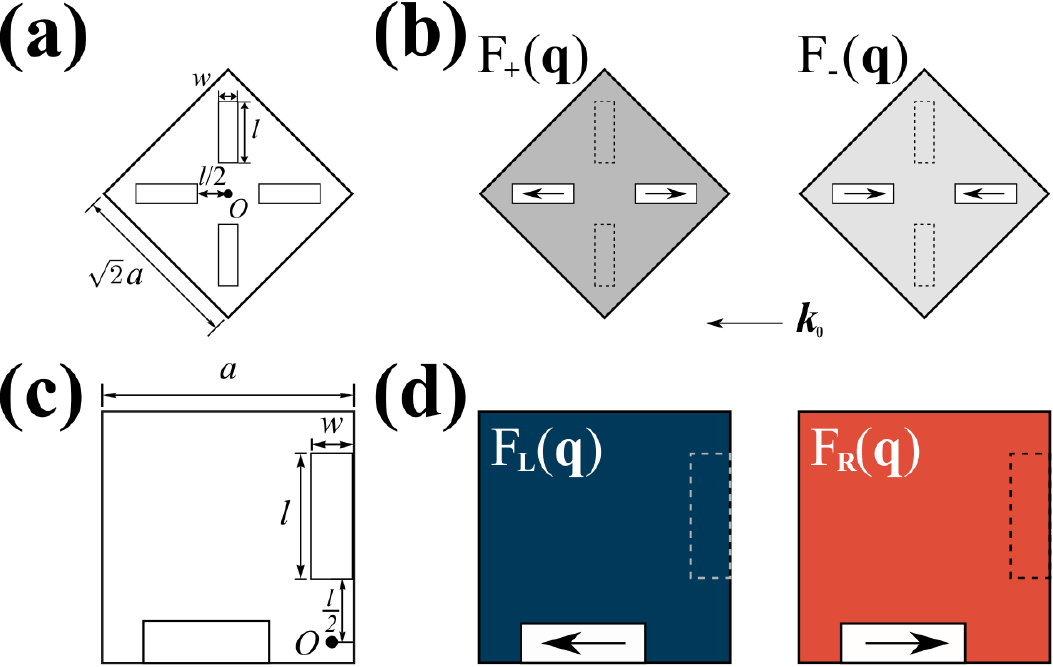}
\caption {
Real space diagrams of the effective magnetic unit cells used in the calculation. 
Only the Ising macrospins the beam is sensitive to ($\vec{k_0}\parallel\vec{M}$) are represented with arrows. 
(a) The dimensions of the effective T1 unit cell and its nanoislands are shown on the left. 
The magnetic motif of each effective unit cell and their magnetic scattering structure factors ($F_+(\mathbf{q})$ and $F_+(\mathbf{q})$) are to the right. 
(b) The dimensions of the effective T2 unit cells and their magnetic scattering structure factors  ($F_L(\mathbf{q})$ and $F_R(\mathbf{q})$). 
The unit cells contain orthogonally oriented Ising nanoislands with sharp-edged rectangular prisms in the structure factor calculation. 
Their lengths and widths are \textit{l} and \textit{w} (height is ignored). 
The lengths of the square T1 and T2 unit cell sides are, respectively, $\sqrt{2}a$ or $a$, where $a$ is the lattice constant.
}  
\end{center}
\label{SFfig}
\end{figure}

Schematics of the bases used in our study are shown in Fig. \ref{SFfig}.
The vector $\boldsymbol{k_0}$ denotes the direction of the incident beam.
The charge basis of a square ASI (Fig. \ref{SFfig}(c)) consists of a pair of two perpendicular nanoislands that are equidistant to a common point, $O$. 
In addition, the long axes of the nanoislands intersect at $O$.
The magnetic basis can differ from this once the Ising macrospins of the nanoislands are taken into account.
The T1 magnetic basis, shown in Fig. \ref{SFfig}(b), is a set of four Ising macrospins that are arranged in a four-fold symmetric pattern around $O$.
The T1 unit cell is accordingly doubled in area and rotated by 45$^{\circ}$ with respect to the charge unit cell.
On the other hand, the T2 basis (Fig. \ref{SFfig}(d)) is a pair of Ising macrospins and its unit cell is identical to the charge unit cell.
Each of these bases have different structure factors.

The magnetic form factors of the nanoislands in the resonant magnetic x-ray scattering process is dependent on their different magnetization directions.
Note that the resonant x-ray scattering process is sensitive to only to the magnetization component collinear to the incident beam, $\boldsymbol{k_0}$, because we used $\boldsymbol{\sigma}$-polarized incident x-rays with low grazing angles ($\sim9^\circ$).\cite{Kortright}
This means that the T1 bases effectively have two Ising macrospins (arrows in Fig. \ref{SFfig}(b)) and the T2 bases effectively have just one Ising macrospin (arrows in Fig. \ref{SFfig}(d)).
Only two basis sets for T2 unit cells and two basis sets for T1 unit cells are used in our calculations.

Calculating a magnetic structure factor of a magnetic basis, $F_{basis}(\boldsymbol{q})$, amounts to calculating the $\boldsymbol{q}$ Fourier component of an unit cell charge density, $\rho_i(\boldsymbol{x})$. 
The sign of this charge density can be positive or negative based on whether the magnetization is parallel or antiparallel to $\boldsymbol{k_0}$:
\renewcommand{\theequation}{\Alph{section}\arabic{equation}}
\begin{equation}
F_{basis}(\boldsymbol{q}) \propto \left( \frac{2\pi}{a} \right)^2 \int_{\text{cell}} e^{i \boldsymbol{q} \cdot \boldsymbol{x} } \boldsymbol{\hat{k}}_0 \cdot \left(\sum_i \boldsymbol{\hat{m}}_i  \rho_i(\boldsymbol{x}) \right) d\boldsymbol{x}.
\label{sofq}
\end{equation}
The sum in the bracket runs over the nanoislands in the unit cell, the $x$-axis is parallel to the incident beam, and the $y$-axis is parallel to the vertical nanoislands shown in Fig. \ref{SFfig}. 
The term $\boldsymbol{\hat{m}}_i$ is the macrospin direction of nanoisland $i$.

We treat the nanoislands as identical rectangular prisms of uniform density $\rho_0$ with the same nominal dimensions of the nanoislands, as specified in Appendix \ref{growth}.
The nanoislands can be defined by a product of unit step functions, $\Theta(x)$: 
\begin{align}
\rho_{T2}(x,y) = \rho_0 \Theta( x - \frac{l}{2} )& \Theta( \frac{3l}{2} - x ) \times \nonumber \\
			&  \Theta( y + \frac{w}{2} ) \Theta( \frac{w}{2} - y  ), \nonumber \\
\rho_{T1,left}(x,y) = \rho_0  \Theta( x - \frac{l}{2} )& \Theta( \frac{3l}{2} - x ) \times \nonumber \\
			&  \Theta( y + \frac{w}{2} ) \Theta( \frac{w}{2} - y  ), \nonumber \\
\rho_{T1,right}(x,y) = \rho_0  \Theta( x + \frac{3l}{2} )& \Theta( - \frac{l}{2} - x ) \times \nonumber \\
			&  \Theta( y + \frac{w}{2} ) \Theta( \frac{w}{2} - y  ).
\label{density-functions}
\end{align}
The subscript indicates whether the function represents one of the T1 nanoislands or the T2 nanoisland shown in Fig. \ref{SFfig}.
 
Using Eq. \ref{sofq} and the density functions of Eq. \ref{density-functions}, the T1 and T2 magnetic structure factors are:
\begin{align*}
F_{+/-}(\boldsymbol{q}) &\propto 
\rho_0 \left ( \frac{ 2 \pi }{ a } \right )^2 \boldsymbol{\hat{k}}_0 \cdot \boldsymbol{\hat{m}}_{left} \times \\
 & \bigg( 
\int_{ \frac{l}{2} }^{ \frac{3l}{2} } e^{ i q_x x } dx -
\int_{ -\frac{3l}{2} }^{ -\frac{l}{2} } e^{ i q_x x } dx 
\bigg) \int_{ -\frac{w}{2} }^{ \frac{w}{2} } e^{ i q_y y } dy,
\\
F_{L/R}(\boldsymbol{q}) &\propto 
\rho_0 \left ( \frac{ 2 \pi }{ a } \right )^2 \boldsymbol{\hat{k}}_0 \cdot \boldsymbol{\hat{m}}
\int_{ \frac{l}{2} }^{ \frac{3l}{2} } e^{ i q_x x } dx 
\int_{ -\frac{w}{2} }^{ \frac{w}{2} } e^{ i q_y y } dy,
\end{align*}
with the index determined by the direction of $\boldsymbol{\hat{m}}$.
(For $F_{+/-}(\boldsymbol{q})$, the relationship $\boldsymbol{\hat{m}}_{left} = -\boldsymbol{\hat{m}}_{right}$ was used.)
Dividing out common factors from the two expressions:
\begin{align*}
F_{+/-}(\boldsymbol{q}) &\propto 
i \boldsymbol{\hat{k}}_0 \cdot \boldsymbol{\hat{m}}_{left}
\sin( l q_x ) \text{sinc}( \frac{l}{2} q_x ) \text{sinc}( \frac{w}{2} q_y ),
\\
F_{L/R}(\boldsymbol{q}) &\propto 
\boldsymbol{\hat{k}}_0 \cdot \boldsymbol{\hat{m}}
e^{i q_x l} \text{sinc}( \frac{l}{2} q_x ) \text{sinc}( \frac{w}{2} q_y ).
\end{align*}
We choose to use the charge/T2 lattice to base our Miller indices on since we found that T2 unit cells play the greatest role in our study. 
Thus, $q_x = 2 \pi \frac{ H }{ a }$ and $q_y = 2 \pi \frac{ K }{ a }$.

These expressions allowed us to calculate the ideal scattered intensity at and around ASI Bragg peak locations in reciprocal space, as described in the text.

\section{Gallery of ASI magnetic textures}
\renewcommand{\thefigure}{\thesection} 

This Appendix illustrates several superdomain arrangements, and their corresponding coherent x-ray magnetic Bragg peak profiles.
We first show several idealized cases that can be used to illustrate how the real-space structure influences the Bragg peak structure.
Then, we show several cusp superdomain scenarios explored before we arrived at the texture shown in the main text.
Lastly, an alternative scenario involving patches of contrasting superdomains distributed in a polarized background is shown.
We explain why the cusp scenario proved to be the better solution than the patch scenario.

\begin{figure}[ht]
\begin{center}
\includegraphics[width=8.75 cm]{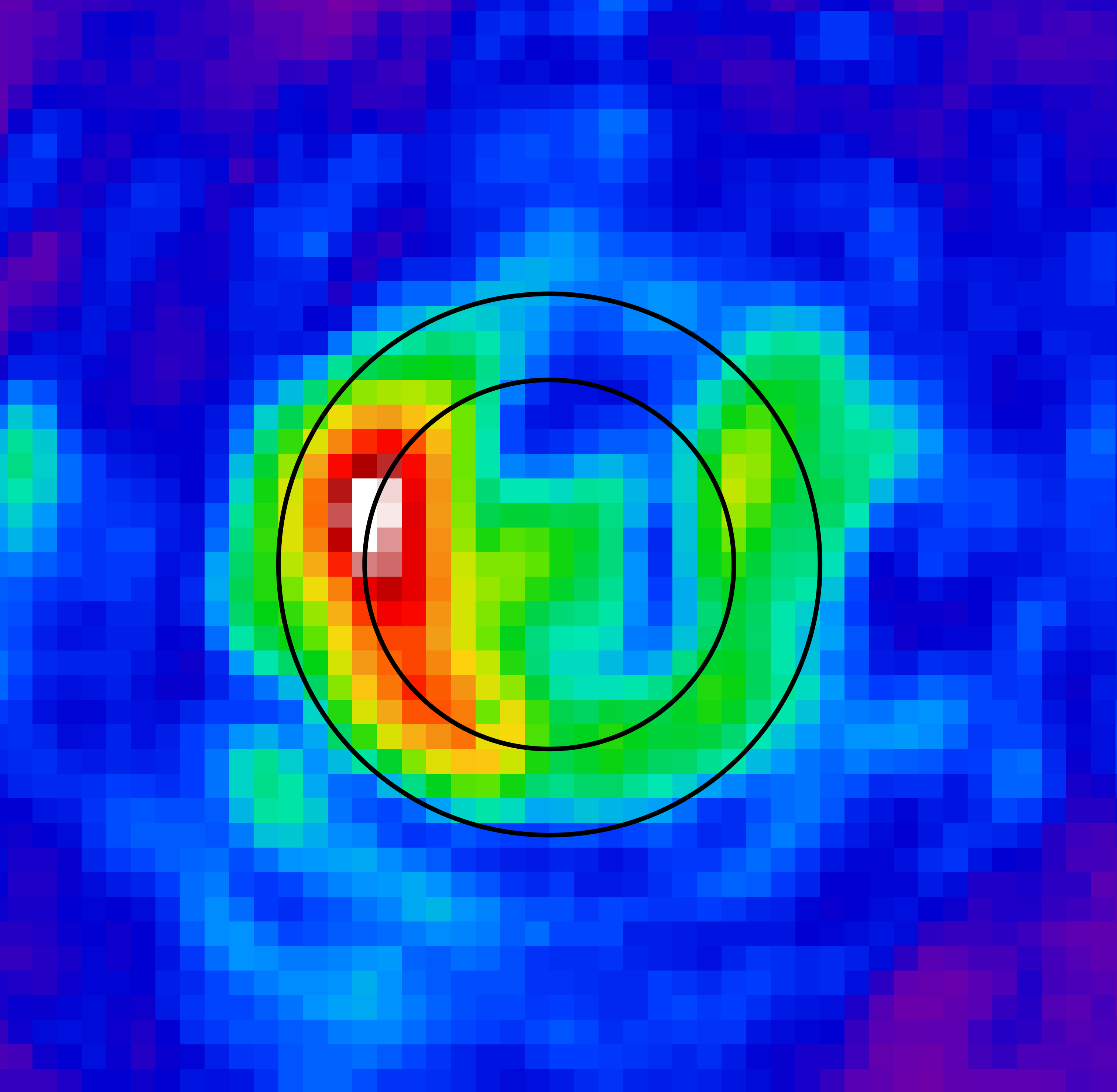}
\caption {
The profile of a split Bragg peak like that shown in the main text, superimposed by concentric black circles that mark the size of the splitting.
}  
\end{center}
\label{dataprofile}
\end{figure}

In nearly every figure that follows, a pair of concentric black circles are superimposed over the calculated pattern.
The circles indicate the size of the Bragg peak splitting seen in the data.
(See Fig. \ref{dataprofile}.)
The best real-space arrangement will cause two lobes of scattering intensity, and a weaker ring of scattering, to form between the circles or on the edge of the inner circle.

	\subsection{Idealized cases}

		\subsubsection{High symmetry textures}
		\setcounter{figure}{0}
		\renewcommand{\thefigure}{\thesection\arabic{figure}} 
The highly symmetric T2 superdomain arrangements shown in Fig. \ref{highsymmetry}, while not realistic, do illustrate how large, contrasting T2 superdomains can cause magnetic Bragg peaks to form complex internal structure.
The Bragg peaks split to form intensity minima at what used to be the center of the Bragg peaks.
Complete destructive interference can occur at the center if the contrasting areas are of equal area.

The splitting occurs along directions in which there is strong magnetic contrast.
For example, in the scenario shown on the left of Fig. \ref{highsymmetry}, the two-fold symmetric checkerboard superdomain structure results in a Bragg peak structure with a similar symmetry.
\begin{figure}[hbt]
\begin{center}
\includegraphics[width=8.75 cm]{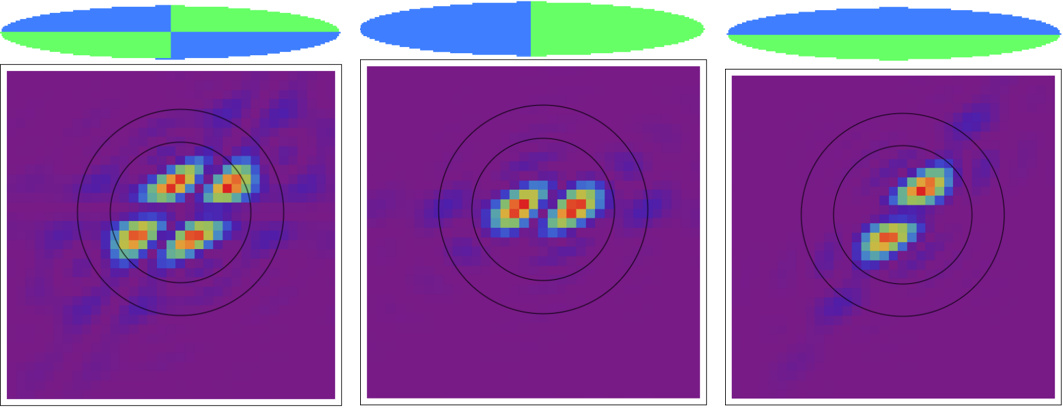}
\caption {
Highly symmetric arrangements of large T2 superdomains and their corresponding coherent Bragg peak profiles. 
The symmetry of the peak splitting reflects the symmetry of the real space arrangements, which are shown above the Bragg peak profiles. 
The concentric black circles indicate the size of the splitting seen in the data. 
}  \label{highsymmetry}
\end{center}
\end{figure}

The fact that the Bragg peak splits to form lobes of intensity to the left and the right of the central Bragg peak position indicates a strong magnetic contrast along the long axis of the beam spot.

		\subsubsection{Linear grating of superdomains}

\begin{figure}[ht]
\begin{center}
\includegraphics[width=8.75 cm]{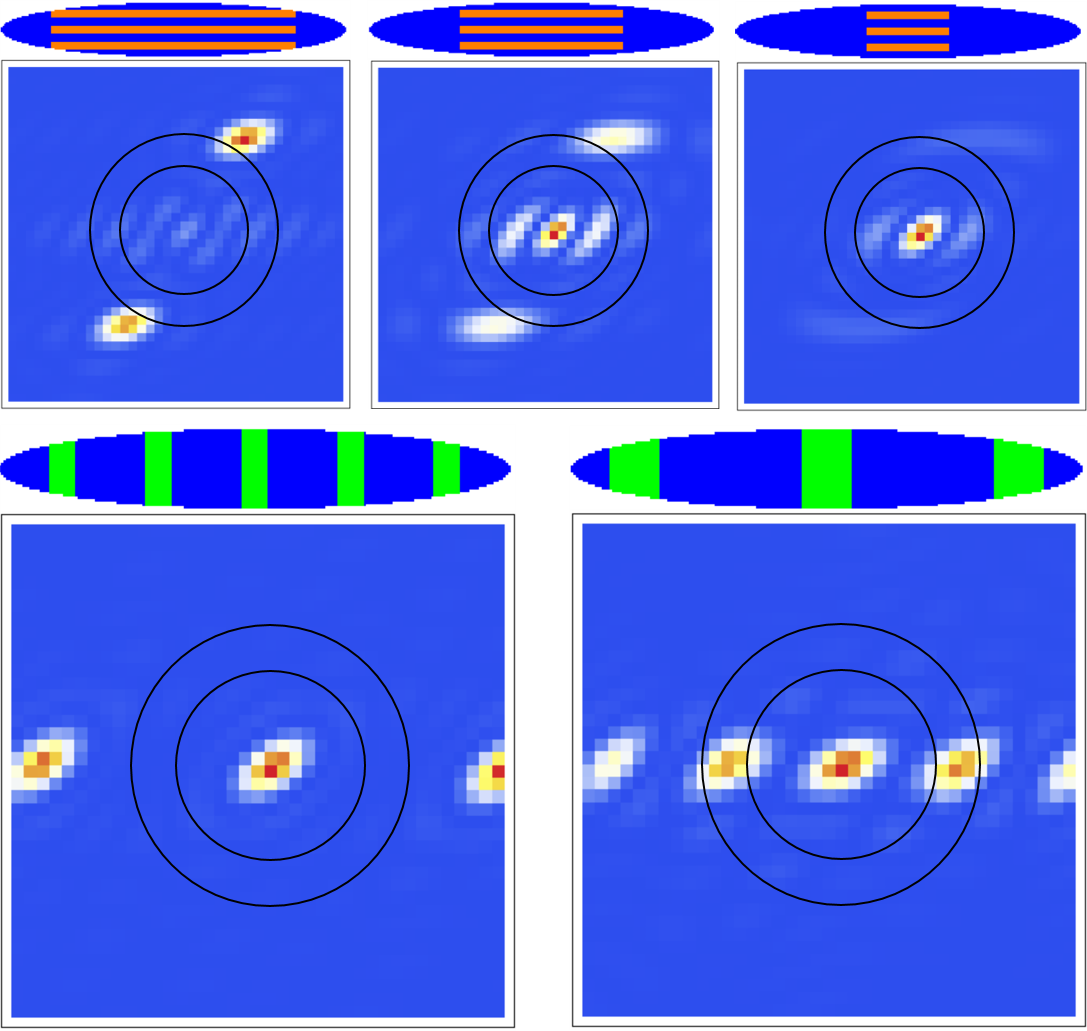}
\caption {
Grating-like arrangements of linear T2 superdomains and their corresponding coherent Bragg peak profiles. 
Satellite peaks form around the Bragg peak position according to the momenta of the grating arrays.
The real space arrangements are shown above the Bragg peak profiles. 
The concentric black circles indicate the size of the splitting seen in the data. 
}  \label{gratings}
\end{center}
\end{figure}

This idealized situation gives an idea of what the Bragg peak structure would be if long, linear T2 superdomains formed in the beam spot.
The situations shown in the top row of Fig. \ref{gratings} are, generally, the type of structure one might expect ifthe superdomains elongated along the direction of the applied magnetic field.
Such T2 superdomains, however, would cause lobes of scattering to form above and below the Bragg peak position, rather than to the left and right of it, as shown in Fig. \ref{dataprofile}.
Based on this, we ruled out such a grating-like structure.

The grating scenarios in the bottom row of Fig. \ref{gratings} do offer clues as to what superdomain length scales would cause the Bragg peak to split as it does in Fig. \ref{dataprofile}.
Specifically, the case with three vertical T2 superdomains, separated by $\sim$25.5$\mu$m or 40\% of the beam spot's length, causes satellite peaks to appear between the two black circles, which indicate the size of the peak splitting.
Therefore, the correct superdomain structure needs to contain this length scale.



	\subsection{Cusps}
\begin{figure}[ht]
\begin{center}
\includegraphics[width=8.75 cm]{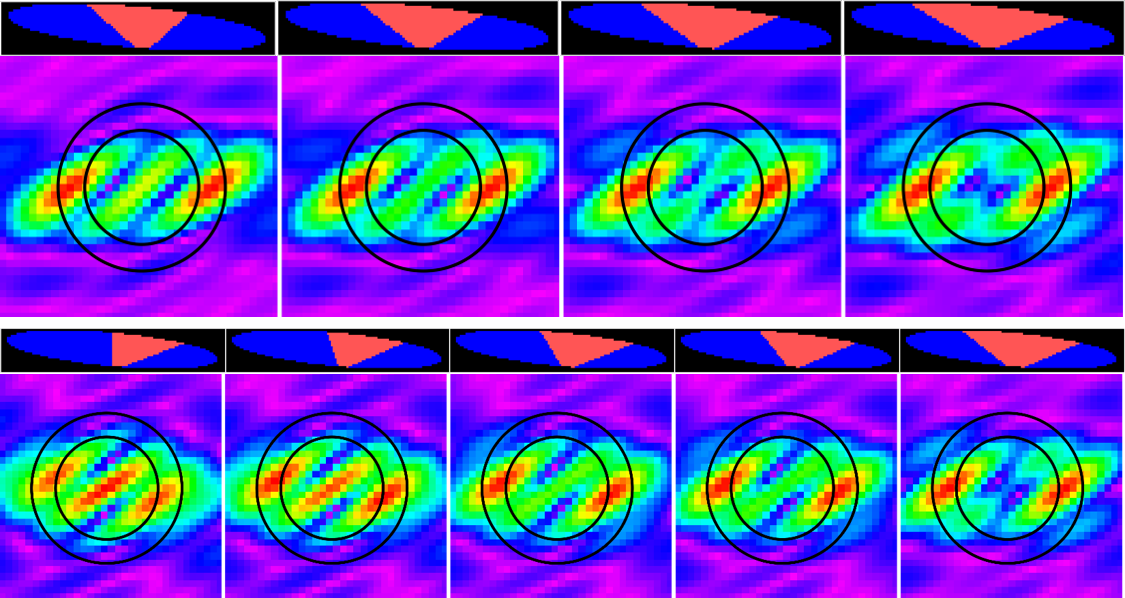}
\caption {
Two scenarios, depicted in the top and bottom rows, in which the opening angle of the T2 superdomain cusp changes. 
In the top row, the opening angle of a cusped superdomain is symmetrically increasing, going from left to right. 
In the bottom row, the opening angle of an asymmetric cusped superdomain is increasing, going from left to right.
The corresponding coherent Bragg peak profiles are shown below the real space arrangements. 
The concentric black circles indicate the size of the splitting seen in the data. 
}  \label{cuspangles}
\end{center}
\end{figure}

We found that a superdomain structure whose boundaries converge to a cusp produced magnetic Bragg peak splitting closest to that observed in the data.
As illustrated in Fig. \ref{cuspangles}, the structures that most closely match the data have nearly equal areas containing parallel and anti-parallel nanomagnets. 
Furthermore, the superdomain structures contain the important length scale illustrated by the grating domains in the previous subsection.

\begin{figure}[ht]
\begin{center}
\includegraphics[width=8.75 cm]{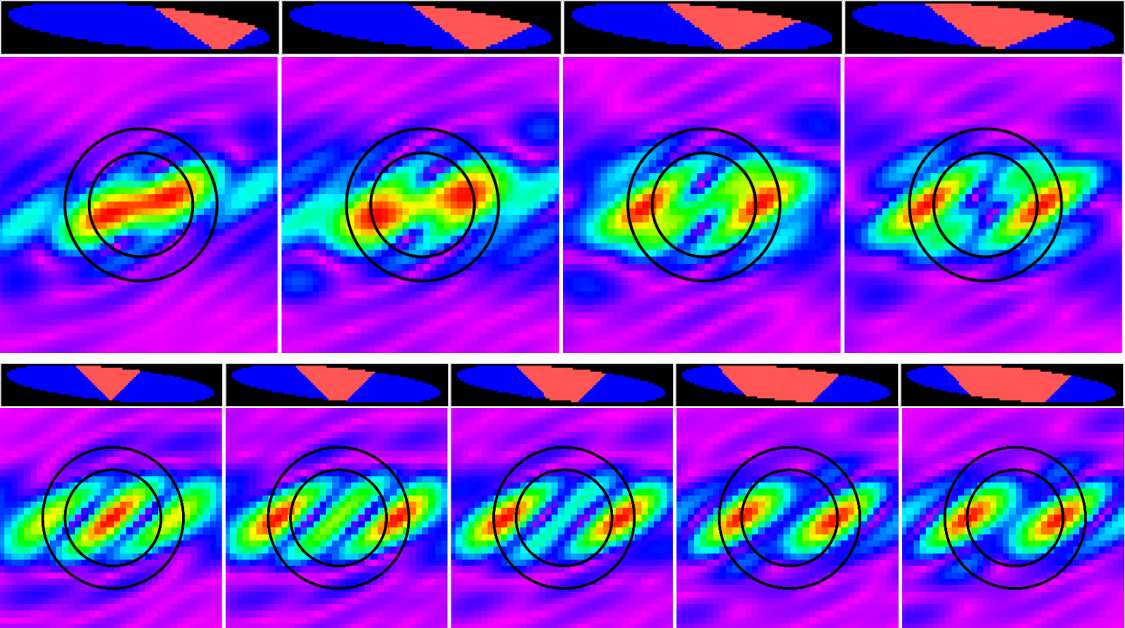}
\caption {
The dependence of the coherent x-ray Bragg peak structure on the position of the cusped T2 superdomain within the beam footprint.
In the top row, the cusped superdomain shifts from right to left.
In the bottom row, a cusped superdomain shifts from top to bottom, such that its cusp is no longer within the beam footprint.
The corresponding coherent Bragg peak profiles are shown below the real space arrangements. 
The concentric black circles indicate the size of the splitting seen in the data. 
}  \label{cuspposition}
\end{center}
\end{figure}

\begin{figure}[ht]
\begin{center}
\includegraphics[width=8.75 cm]{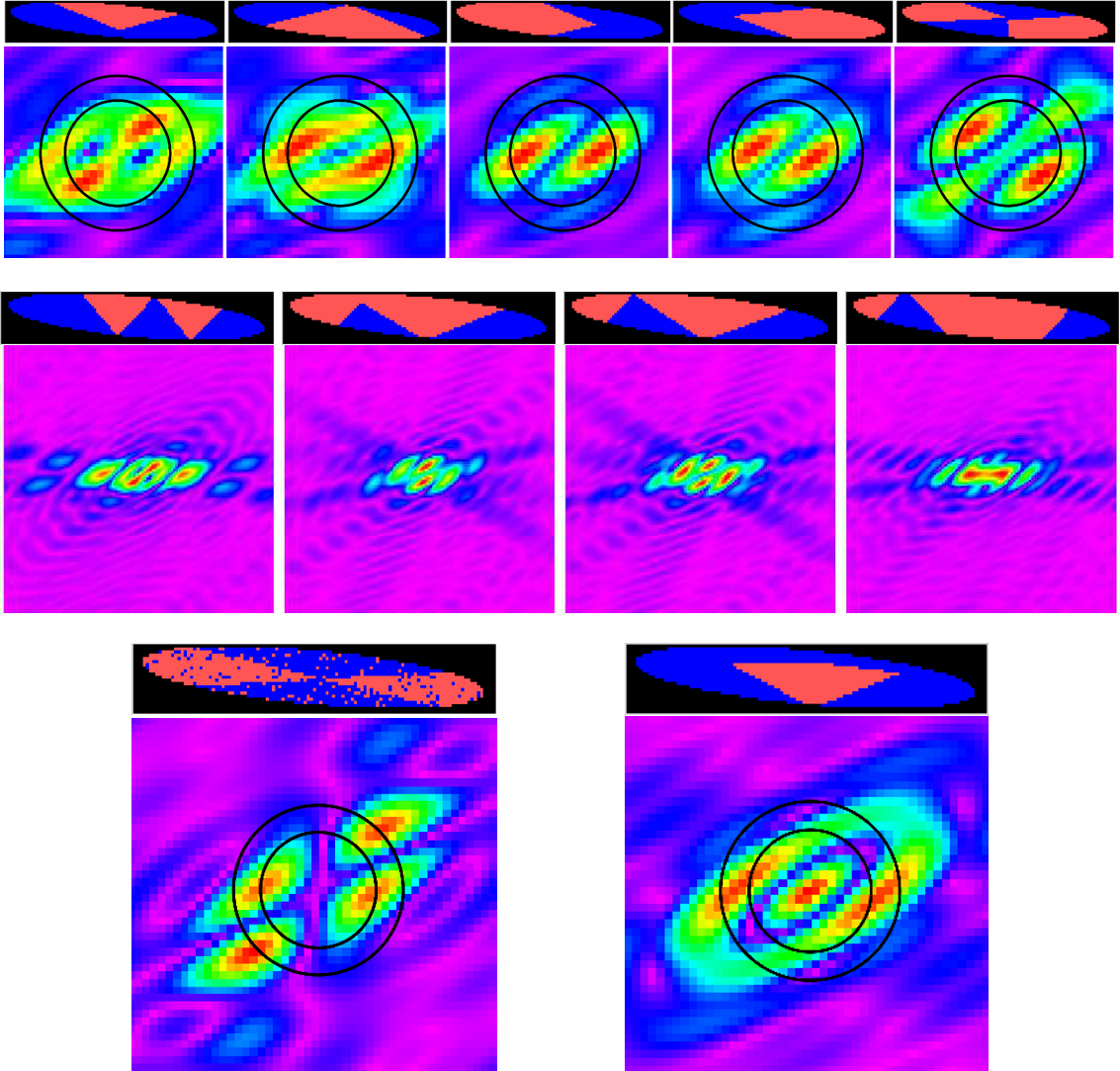}
\caption {
Other cases in which T2 superdomains display cusps and their corresponding coherent x-ray Bragg peaks.
The middle row of panels shows a wider area of reciprocal space around a Bragg peak than is shown in either the top or bottom rows, as was necessary to show the scattering that extends far from the Bragg peak position.
}  \label{cuspother}
\end{center}
\end{figure}

In search of the most plausible superdomain structure, we considered many cusped domain structures, some of which are shown in Figures \ref{cuspangles}-\ref{cuspother}. 
In Fig. \ref{cuspangles}, cusped superdomains of various shapes are shown, along with their Bragg peak structures. 
The top row demonstrates that the angle of the cusp must be such that nearly equal areas of parallel and anti-parallel nanomagnets are present.
Superdomains that are nearly symmetric about the vertical axis also concentrate the scattering intensity in the lobes to the left and right of the Bragg peak center.

Figure \ref{cuspposition} illustrates the effects of placing the cusped superdomain away from the center of the illuminated area.
In the top row, the superdomain from being right of center to nearly centered.
The most off-center case causes the two lobes to merge near the Bragg peak center, which separate as the superdomain becomes more centered.
The bottom row shows what it might look like if the superdomain were shifted down, highlighting the role of the cusp in forming a split Bragg peak.
In this case, the superdomain is still laterally centered and, thus, the two lobes of scattering remain and are nearly at the right positions in all of the five cases.
Comparing this bottom row of figures to the top row of Fig. \ref{cuspangles}, the cusp and the angled superdomain boundaries cause scattered intensity to form a weak ring shape around the black circles shown in the figures, closely matching the Bragg peak splitting seen in the data.
Therefore, the cusp of the superdomain plays an important role in creating a ring of scattering, rather than a peak that is split into two parts.

The mere presence of a cusp is not sufficient to create the right magnetic Brag peak splitting, however. 
As shown in Fig. \ref{cuspother}, the orientation of the superdomain is important to capturing the size of the peak splitting.
Furthermore, the superdomain must completely span the beam spot.
The top row, left panel of Fig. \ref{cuspother} shows a case when there is a connection between the blue areas, which are oppositely polarized to the red cusp superdomain. 
This connection shifts the scattering intensity away from the peaks to the left and right of the Bragg peak center.
Another case like this is shown in the bottom row, right panel, in which a wedge-shaped superdomain is roughly in the middle of the beam spot. 
While there is strong scattering to the right and left of the Bragg peak position, and the size of the split is right, this case is discounted because it creates scattering well away from the black circles.
Cases with multiple cusps also cause scattering outside of the outer black circle, which also disqualify them as likely structures.

\begin{figure}[hbt]
\begin{center}
\includegraphics[width=8.75 cm]{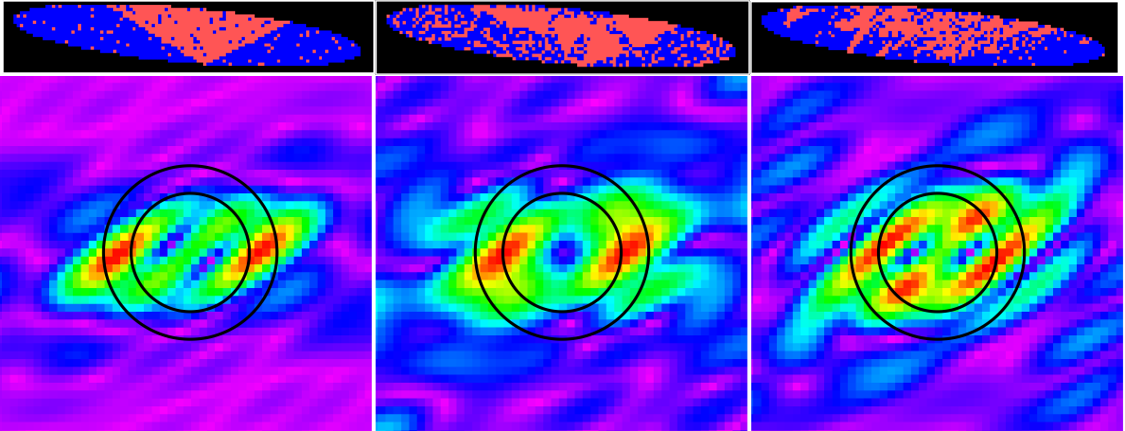}
\caption {
The best cusp configuration is subjected to different degrees of disorder.
In each case, the size of the Bragg peak splitting remains the same.
}  \label{cuspdisorder}
\end{center}
\end{figure}

Figure \ref{cuspdisorder} illustrates the effect of incompletely polarized superdomains, and other forms of disorder, on the scattering pattern. 
Even when the boundary between the cusped superdomain and the magnetically contrasting background is poorly defined, and many oppositely polarized areas exist in the cusped superdomain, the scattered intensity is still distributed within the black circles.
The size of the peak splitting and ring-like scattering pattern in the central figure of Fig. \ref{cuspdisorder} most closely resembles the data shown in Fig. \ref{dataprofile}.
It is this calculation, after a convolution with a gaussian function to simulate the effect of a partially coherent incident beam, that we show in Fig. 4(a) of the main text.

	\subsection{Alternative case: patches}

\begin{figure}[ht]
\begin{center}
\includegraphics[width=8.75 cm]{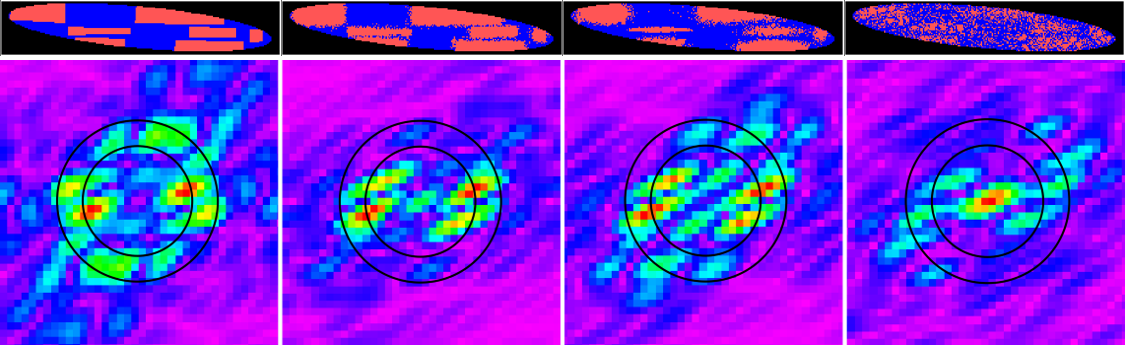}
\caption {
Four scenarios in which large T2 superdomains are arranged over the illuminated area.
The two lobes of intensity are replicated, though the size of the splitting between those lobes is too small.
The Bragg peak structures in these scenarios are quite sensitive to disorder.
}  \label{patches}
\end{center}
\end{figure}

Another set of superdomain structures were also seriously considered.
In these cases, patches of superdomains are distributed around the beam spot.
The left panel of Fig. \ref{patches} shows a superdomain configuration that gives rise to a scattering ring of the correct size and intensity distribution.
However, small variations of the patches (\textit{e.g.}, position, size, disorder) cause the magnetic Bragg peak structure to dramatically change.
This instability against small changes in superdomain configuration, occur in the ASI as it is subjected to one magnetic field cycle to another, is the main reason we abandoned further investigation of these structures.

\bibliography{bibliography}

\end{document}